\documentclass{article}

\usepackage{epsfig }

\begin{document}

\title{Impact of saturation on spin effects\\  in proton-proton
                    scattering\footnote{presented by O.V.S.
at the Advanced Studies Institute ``Symetries and Spin" 
(SPIN-Praha-2004), Prague, July 5 - July 10, 2004.} }

\author{O.\,V. Selyugin\footnote{
JINR, Bogoliubov Laboratory of Theoretical Physics, 141980 Dubna, Russia, e-mail: selugin@thsun1.jinr.ru} ~~and
~~J.-R. Cudell\footnote{Institut de Physique, B\^at. B5a, Universit\'e de Li\`ege, Sart Tilman, B4000  Li\`ege, Belgium, e-mail: JR.Cudell@ulg.ac.be}}

\maketitle

\begin{abstract}

For pomerons described by a sum of two simple-pole terms, a soft
and a hard pomeron, the unitarity  bounds from saturation in
impact-parameter space are examined.  We consider the effect of
these bounds on observables linked with polarisation, such as the
analyzing power $A_N$ in elastic proton-proton scattering, for LHC
energies.  We obtain the $s$ and $t$
dependence of the Coulomb-nuclear interference at small momentum
transfer, and show that the effect of the hard pomeron may be
observed at the LHC.
\end{abstract}

\section{Introduction}
The study of elastic scattering requires a detailed knowledge of
the properties of the pomeron, which is the dominant interaction
of hadrons at high energies. In this case, the structure and spin
properties both of the hadron and of the pomeron play a special
role.

There are two regimes for the pomeron, the ``soft"
non-perturbative pomeron, and the perturbative-QCD ``hard"
pomeron. The soft pomeron dominates in high-energy hadron-hadron
diffractive reactions at long distances while the hard pomeron
dominates in high-energy short-distance scattering \cite{bj} and
determines the very-small-$x$ behaviour of deep inelastic
structure functions and spin-averaged gluon distributions.

The ``soft" pomeron of the standard form with
$ \alpha_{pom}(0)=1 + \epsilon$.
was introduced in \cite{lan1}%
 The observed growth of inelastic cross sections and the
multiplicity coincide with these idea.
The perturbative QCD leading log calculation of the gluon ladder diagrams
gives the following result \cite{kur}
\begin{eqnarray}
   \epsilon = 12 \  \alpha_{s}/ \pi \  \ln{2} \sim 0.5  .
\end{eqnarray}
 Really,
the new global QCD analysis of data for various hard scattering processes
leads to the small x behaviour of the gluon structure function
 determined by the hard pomeron contribution   \cite{lai}
$            g(x) \sim 1/x^{1+\epsilon} $
with $\epsilon = 0.3$.
A recent analysis \cite{mrt} of the experimental data for total
cross sections and $\rho$-parameters of hadron-hadron and
photon-hadron scattering reveals the possibility that both soft
and hard pomerons are simple poles, and that both contribute to
soft physics: if the soft pomeron is a simple pole, the hard
pomeron is in fact needed to describe simultaneously the energy
dependence of the total cross sections (hence the imaginary part
of the scattering amplitude at $t=0$) simultaneously with the
value of $\rho(s, t=0)$---the ratio of the real to the imaginary
part of the hadron elastic spin-non-flip scattering
amplitude---(hence the real part of this amplitude).

Note that the contribution of the hard pomeron to the elastic
amplitude leads to the saturation of the unitarity bound in the
profile function at some impact-parameter values at very high
energies.

Here, by ``saturation", we mean, in the $S$-matrix language, that
for some distance between scattering particles, the maximum
possible scattering is reached. This distance is correlated with
the angular momentum $l$ and part of the scattering reaches the
black disc limit (BDL).

It is that last meaning of ``saturation", which is directly
connected with the unitary property of the scattering amplitude,
that we shall use in this work.


 \section{The elastic-scattering amplitude at high energy}
The presence of a hard pomeron in elastic diffractive processes
with a large intercept $1+\epsilon_2 =1.45$ \cite{mrt,lnd-hp} will
lead to a lowering of the energy where saturation starts: it now
appears as we approach the TeV region.
It is not obvious
how the total cross sections will grow with energy after
partial saturation, especially in the energy region of LHC.

  One can show that saturation leads to a behaviour of the total
cross sections at LHC energy which is almost model-independent.
Indeed, different models can be obtained by varying the the
profile function of the hadrons. The BDL leads to a cut of this
function when it reaches 1. If we take a sharp cut with a break in
the derivative, then we obtain strong edge effects from the disk
corresponding to the radius of saturation $b_s$.  To remove such
nonphysical behaviour, we input an additional function which
provides a smooth matching between the central black disk and the
rest of the amplitude, and consider various profile functions
(exponential, Gaussian, multipole).

We find in fact that the energy dependence of the imaginary part
of the amplitude and hence of the total cross section depends on
the form of $f(b)$, {i.e.} on the $s$ and $t$ dependence of the
slope of the elastic scattering amplitude, but that fitting these
quantities to existing data removes most of the uncertainty. It is
unlikely that the LHC will help us choose the right profile
function, or let us decide whether saturation has been reached.
So, we need additional information.

Therefore, we investigate the impact of the hard pomeron on the
polarisation of the elastic proton-proton scattering at LHC
energies and at small momentum transfer. One of the most important
effects is the Coulomb-Nuclear Interference at small
momentum transfer
 \cite{schwinger,lap,lead},
which mostly comes from the interference of the imaginary part of
the hadron spin-non flip amplitude with the electromagnetic part
of the spin-flip amplitude.

The differential cross section and analysing power $A_N$  are
defined as follows:
\begin{eqnarray}
\frac{d\sigma}{dt}&=& \frac{2
\pi}{s^2}(|\Phi_1|^2+|\Phi_2|^2+|\Phi_3|^2
  +|\Phi_4|^2+4|\Phi_5|^2), \label{dsth}\\
  A_N\frac{d\sigma}{dt}&=& -\frac{4\pi}{s^2}
                 Im[(\Phi_1+\Phi_2+\Phi_3-\Phi_4) \Phi_5^{*})],  \label{anth}
\end{eqnarray}
in terms of the usual helicity amplitudes $\Phi_i$.

In the general case, the  form of $A_N$ and the position of its
maximum depend on the parameters of the elastic scattering
amplitude: $\sigma_{\rm tot}$,  $\rho(s,t)$, the Coulomb-nuclear
interference phase  $\varphi_{\rm cn}(s,t)$ and the elastic slope
$B(s,t)$. For the definition of new effects at small angles, and
especially in the region of the diffraction minimum, one must know
the effects of the Coulomb-nuclear interference with sufficiently
high accuracy. The Coulomb-nuclear phase was calculated in the
entire diffraction domain taking into account  the form factors of
the  nucleons \cite{prd-sum}.

The total helicity amplitudes can be written as
\begin{eqnarray}
  \Phi_i(s,t) = \phi^h_{i}(s,t)
        + \phi_{i}^{\rm em}(t) \exp[i \alpha_{\rm em} \varphi_{\rm cn}(s,t)],
 \end{eqnarray}
where $\alpha_{\rm em}=1/137$ is the electromagnetic constant,
$\phi^h_{i}(s,t)$ describes the strong interaction of hadrons,
$\phi_{i}^{\rm em}(t)$  their electromagnetic interaction, and
$\varphi_{\rm cn}(s,t)$ is the electromagnetic-strong interference
phase factor. Therefore, to determine the hadron spin-flip
amplitude at small angles, one should take into account all
electromagnetic and Coulomb-nuclear interference (CNI) effects.

In this paper, we define the hadronic and electromagnetic
spin-non-flip amplitudes as
\begin{eqnarray}
  F^{h}_{\rm nf}(s,t)
   &=&
    = \frac{1}{2s}\left[\phi^h_{1}(s,t) + \phi^h_{3}(s,t)\right]; \ \
      \\
 F^{c}_{\rm nf}(s,t)
    &=&
  = \frac{1}{2s}\left[\phi^{\rm em}_{1}(s,t) + \phi^{\rm em}_{3}(s,t)\right].
    \end{eqnarray}
Taking into account the Coulomb-nuclear phase $\varphi_{\rm cn}$,
we obtain
\begin{equation}
Im F_{\rm nf}^{c} = \alpha_{\rm em} \varphi_{\rm cn} F_{\rm
nf}^{c} .
\end{equation}

Let us now examine the behaviour of the analysing power
(\ref{anth}),  which can be rewritten as
\begin{eqnarray}
  \frac{A_N}{2}\frac{d\sigma}{dt}& =&
  (ImF_{\rm nf}ReF_{\rm sf}-ReF_{\rm nf}ImF_{\rm sf}) \\ \nonumber
 &=&  [( ImF_{\rm nf}^h ReF_{\rm sf}^c +ImF_{\rm nf}^c ReF_{\rm sf}^c
        - ReF_{\rm nf}^h ImF_{\rm sf}^c
   -ReF_{\rm nf}^c ImF_{\rm sf}^c )\\ \nonumber
     &+& (ImF_{\rm nf}^h ReF_{\rm sf}^h -ReF_{\rm nf}^c ImF_{\rm sf}^h
        +  ImF_{\rm nf}^c ReF_{\rm sf}^{h} -Re F_{\rm nf}^{h} Im F_{\rm sf}^{h})]. \label{anf}
\end{eqnarray}

Equation (\ref{anth}) was applied at high energy and at small
momentum transfer, with the following usual assumptions
for hadron spin-flip amplitudes:\\
$\bullet$ $\phi_{1}=\phi_{3}$, $\phi_{2}=\phi_{4} = 0$, \\
$\bullet$ the slopes of the hadronic spin-flip and spin-non-flip
amplitudes are equal.

In this kinematic region, the analysing power can then be
rewritten as
\begin{eqnarray}
\frac{A_N}{2}\frac{d\sigma}{dt}& =& -(ImF_{\rm nf}^h  ReF_{\rm
sf}^c + Im F^{c}_{\rm nf} Re F^{c}_{\rm sf})
    + ImF_{\rm sf}^c  (Re F_{\rm nf}^{c} + Re F_{\rm nf}^{h})
\end{eqnarray}

   If we know the parameters of the hadron spin non-flip amplitude,
 the measurement of the analyzing power at small transfer momenta
 helps us to find the structure
 of the hadron spin-flip amplitude.

\subsection{Soft and hard pomeron as simple poles}

Let us extend the Donnachie-Landshoff (DL) \cite{DL} approach as
it is a well-known model for hadron-hadron  elastic scattering. A
modern analysis of the existing experimental data for the
hadron-hadron and photon-hadron total cross sections and
$\rho(s,t=0)$ parameters suggests a possible contribution of the
hard pomeron \cite{mrt,lnd-hp}.

In the DL approach, the $pp$-elastic scattering amplitude is
proportional to the hadrons form-factors and can be approximated
at small $t$  by \cite{lnd-hp}:
\begin{eqnarray}
T(s,t) \ =  [\ h_{1} \ (s/s_0)^{\epsilon_1}
    e^{\alpha^{\prime}_1 \  t \ \ln (s/s_0)}
   \ + \    \ h_{2} \ (s/s_0)^{\epsilon_2}
             e^{\alpha^{\prime}_2 \  t \ \ln (s/s_0)} ]
   \ F^2(t).
\end{eqnarray}
where $h_1=4.47$  and $h_2 = 0.005$ are the coupling of the
``soft'' and ``hard'' pomerons, and $1+\epsilon_1 =1.073$,
$\alpha^{\prime}_1=0.3$\,GeV$^2$, and  $1+\epsilon_2=1.45$,
$\alpha^{\prime}_2=0.10$\,GeV$^2$ are the intercepts and the
slopes of the two pomeron trajectories. The normalisation $s_0$
will be set to $1$\,GeV$^2$ below and $s$ implicitly contains the
phase factor $\exp(-i \pi/2)$. $F^2(t)$ is  the square of the
Dirac elastic form factor,

 \begin{eqnarray}
  F^2(t)=\frac{4 m_p^2-2.79 t}{4 m_p^2-t} \frac{1}{1-t/\Lambda^2}
                                     \label{dff}
 \end{eqnarray}
where $m_p$ is the mass of the proton and
$\Lambda^2=0.71$\,GeV$^2$. It can be approximated by the sum of
three exponential \cite{book}:
\begin{eqnarray}
 F^2(t)\approx f_a e^{d_a \ t}  +   f_b e^{d_b \ t}
 + f_c e^{d_c \ t}.   \label{eff}
 \end{eqnarray}
with
 $$f_a=0.55 ,  \ \ \ f_b=0.25,  \ \ \ f_c=0.20,$$
  and
$$d_a=5.5 \ {\rm GeV}^{-2}, \ \ \  d_b=4.1 \ {\rm GeV}^{-2}, \ \ \ d_c=1.2 \ {\rm GeV}^{-2} .$$

We then obtain in the impact parameter representation a specific
form for the profile function $\Gamma(b,s)$ \cite{dif04}. One can
see that at some energy and at small $b$,  $\Gamma(b,s)$ reaches
the black disk limit.
For one-(soft+hard)-pomeron exchange 
this will be in the region $\sqrt{s} \approx 1.5  $\,TeV. If one
adds to the model the cut due to 2-pomeron exchanges \cite{DL},
the resulting $\Gamma_2$ will saturate at $\sqrt{s}=4.5  $\,TeV.

From the DL model, we obtain in the impact parameter
representation the following form of  the profile function
$\Gamma(b,s)$:
\begin{eqnarray}
 \Gamma(b,s)  &=&
   h_{1a} \exp(-\frac{x^2}{r^2_{1a}})
    +  h_{1b} \exp(-\frac{x^2}{r^2_{1b}})  +  h_{1c} \exp(-\frac{x^2}{r^2_{1c}})
                                                    \nonumber \\
     &+&  h_{2a} \exp(-\frac{x^2}{r^2_{2a}})
       +  h_{2b} \exp(-\frac{x^2}{r^2_{2b}}) + h_{2c} \exp(-\frac{x^2}{r^2_{2c}}),  \label{pf}
\end{eqnarray}
where the variables $h_i,\ i=a, b, c$ are
\begin{eqnarray}
         h_{1i}
 = 2  {f_i}  h_1  s^{\epsilon_1} / r^2_{1i}; \ \  \ 
         h_{2i} =
2  {f_i}  h_2  s^{\epsilon_2} / r^2_{2i}.
\end{eqnarray}
with
\begin{eqnarray}
r_{1i}^2 = 4  \left(d_i + \alpha^{\prime}_1
\log({s}/{s_0})\right); \ \ \ r_{2i}^2 = 4  \left(d_i +
\alpha^{\prime}_2 \log({s}/{s_0})\right) .
\end{eqnarray}

 Here the value $s$ contains the phase factor $\exp(-i \pi/2)$.
Saturation of the profile function will control the behaviour of
$\sigma_{\rm tot}$ at super-high energies.
We implement saturation as explained above, by using a smooth
matching between the saturated region ($ \Gamma(b,s)=1$, $b<b_s$)
and the unsaturated one ($ \Gamma(b,s)$ given by (\ref{pf}).
\begin{figure}[!ht]
~\vglue -1.5cm
\centerline{\mbox{\epsfysize=80mm\epsffile{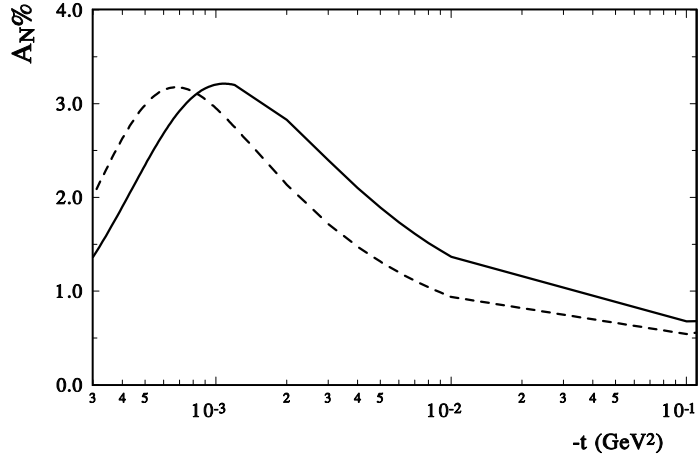}}}\vglue
-0.6cm
 \caption{ $A_N$ due to the interference of electromagnetic
and strong amplitudes, calculated at $\sqrt{s}  =  20 $\,TeV.
(hard line --- with saturation, with contributions from the soft
and the hard pomerons; dashed line  ---  without saturation, with
a contribution only the soft pomeron, normalised to the total
cross section at $\sqrt{s}  =  550$\,GeV). } \label{Fig_1}
\end{figure}

Let us now calculate the CNI effect at LHC energies in different
approaches. On the one hand, the contribution of the hard pomeron
will lead to a growth of the real part of the spin-non-flip
elastic scattering amplitude, but on the other hand, the
contribution of the hard pomeron in the profile function at small
impact parameters leads to the saturation of the unitarity bound,
and to a decrease of the real part of the hadron spin-non-flip
amplitude.  The calculated $A_N$, which comes from the CNI-effect,
in the framework of the Donnachie--Landshoff model with a
contribution from the hard pomeron is shown in Fig.~1.

In this figure, the hard line represents the result from
saturation with contributions from the soft and from the hard
pomerons whereas  the dashed line is calculated without saturation
but with only a soft pomeron. Of course the predictions for LHC
energies depend on the values of the couplings and intercepts of
the pomeron(s). For both cases, we choose these parameters to
obtain a correct description of the total cross sections at high
energies ($\sqrt{s} > 50$\,GeV). From Fig.~1, we can see that the
shape of the analysing power is practically the same in both
cases.

In Fig. 2a, we show  the energy dependence of the value of the
maximum of  $A_N$  for three cases. In the first one, we use
saturation and contributions from  the soft and the hard pomerons.
In the second one, we do not use a unitarity/saturation bound and
we allow a free growth of the profile function.

We then obtain a decrease of the maximum CNI-effect faster than in
the case of the saturation regime but only at  very high energies
($\sqrt{s} > 10 $\,TeV). This means that saturation tampers the
growth of the real part which comes from the hard pomeron.
Finally, the third case shows the standard behaviour of of the CNI
effect when we take into account only the contribution of the soft
pomeron (with intercept 1.07), which reaches the unitarity bound
only at super-high energies beyond the LHC. At the LHC, the value
of the maximum of $A_N$ changes slowly with the energy.

In Fig. 2b, we show  the energy dependence of the value of the
maximum of $A_N$  in these three cases. It increases faster in the
case of the saturation regime especially at sufficiently large
energies, whereas the smallest change corresponds to the standard
analysis with a contribution from the soft pomeron only.
\begin{figure}\begin{center}
~\vglue -1.2cm \vbox{\mbox{\epsfxsize=110mm\epsffile{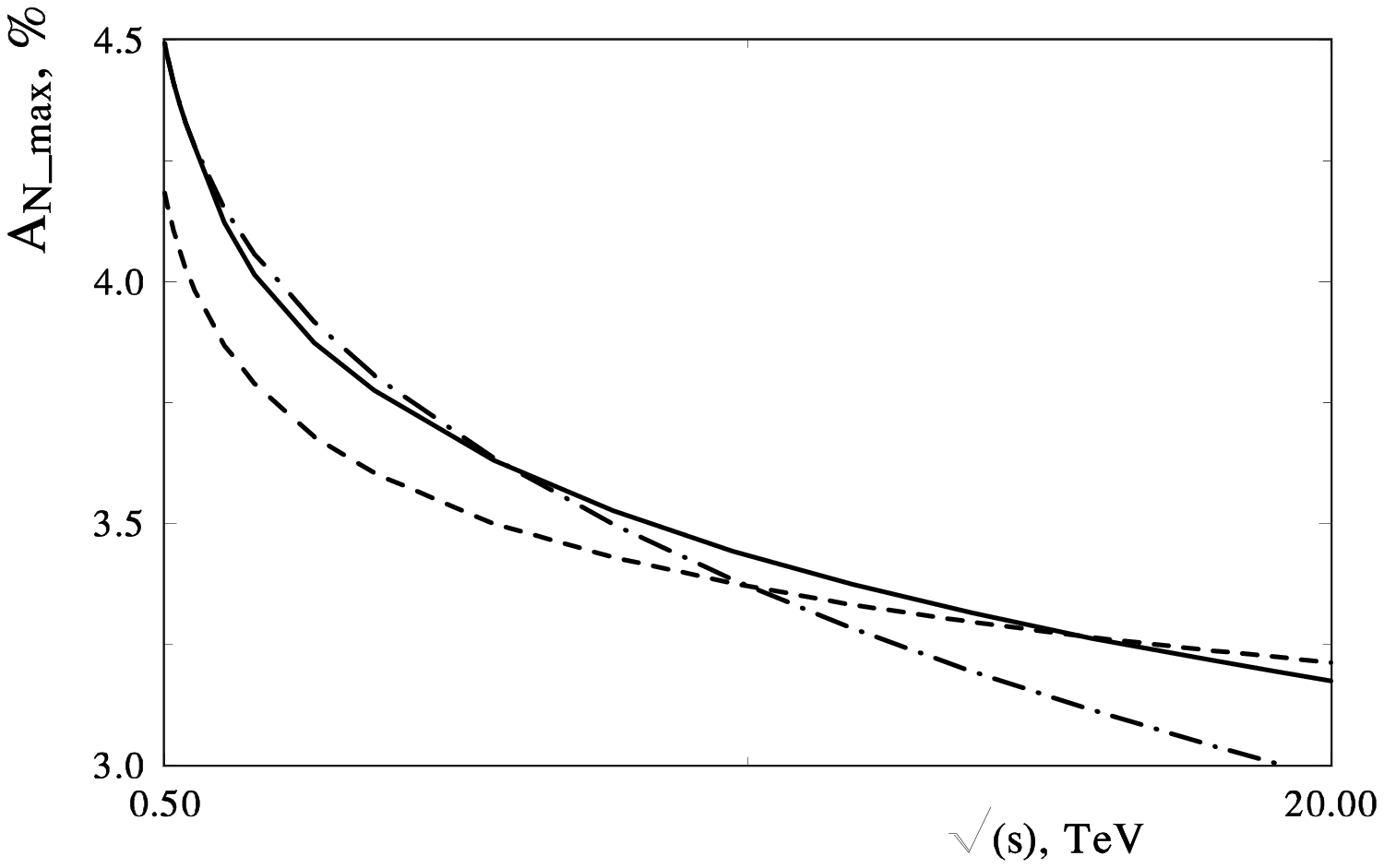}}~\vglue -4.5cm\hglue 4.5cm (a)\\[2.5cm]
\mbox{\epsfxsize=110mm\epsffile{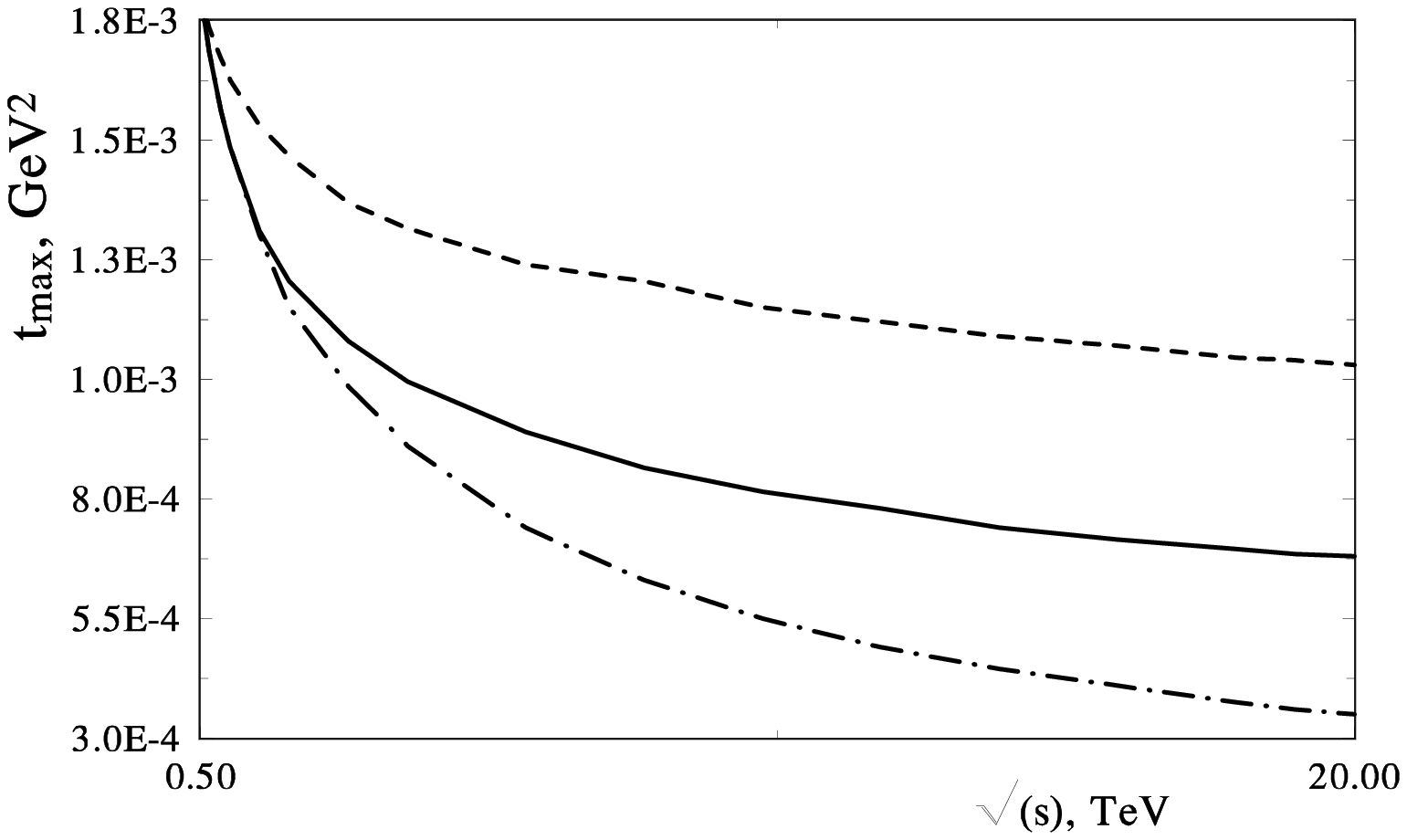}}\\
~\vglue -5.5cm\hglue 4.5cm (b)\vglue +4cm }
\end{center}\caption{ The energy
dependence of a) the maximum value of $A_N$ and b) the position in
$t$ of the maximum of $A_N$ (hard line: in the saturation regime,
with contributions from the soft and the hard pomeron; the
dash-dotted line: the same, but without saturation; dashed line:
without saturation, with a contribution from the soft pomeron
only, normalised to the total cross section at $ \sqrt{s} \ = \
550 $\,GeV). }
\end{figure}

\section{Conclusion}
In the presence of a hard pomeron, saturation effects can change
the behaviour of some characteristic features of the diffractive
scattering amplitudes at the LHC.
However, accurate measurements of the analysing power in the
Coulomb-nuclear interference region can map the structure of the
hadron spin-flip amplitude, and this will give us further
information about the behaviour of hadronic interactions at large
distances.
Large-distance spin-flip contributions can be taken into account,
for example in the peripheral dynamic model \cite{zpc,yaf-str}.
Saturation will lead to a relative growth of the contribution of
peripheral interactions, and to changes in the energy dependence
of the differential cross sections  at moderate momentum transfer.
This is especially true in the case of the energy dependence of
the Coulomb-nuclear interference at small $t$. Such saturation
effects can in principle be observed at the LHC.\bigskip

\noindent {{\bf Acknowledgements:}
The authors would like to thank P.V. Landshoff for helpful
discussion. This research was conducted while O.V.S. was a
Visiting Fellow of the FNRS, Belgium.}


 \end{document}